\documentclass[11pt]{iopart}
\usepackage{graphicx}
\usepackage{color} 
\usepackage{transparent}
\usepackage{iopams}  
\usepackage[normalem]{ulem}

\begin{document}
\article[Experimental study of the role of trap symmetry...]{}{Experimental study of the role of trap symmetry in an atom-chip interferometer above the Bose-Einstein condensation threshold}

\author{M. Dupont-Nivet$^{1}$\footnote{Corresponding author: matthieu.dupontnivet@thalesgroup.com}, R. Demur$^{1}$, C. I. Westbrook$^{2}$ and S. Schwartz$^{3}$}

\address{${}^{1}$Thales Research and Technology France, 1 av. Augustin Fresnel, 91767 Palaiseau, France}
\address{${}^{2}$Laboratoire Charles Fabry de l'Institut d'Optique, 2 av. Augustin Fresnel, 91127 Palaiseau, France}
\address{${}^{3}$Laboratoire Kastler Brossel de l'Ecole Normale Sup\'erieure, 24 rue Lhomond, 75231 Paris Cedex 05, France}

\begin{abstract}
{We report the experimental study of an atom-chip interferometer using ultracold rubidium 87 atoms above the Bose-Einstein condensation threshold. The observed dependence of the contrast decay time with temperature and with the degree of symmetry of the traps during the interferometer sequence is in good agreement with theoretical predictions published in [Dupont-Nivet \emph{et al.}, NJP 18, 113012 (2016)]. These results pave the way for precision measurements with trapped thermal atoms. } 
\end{abstract}

\vspace{2pc}

\maketitle

\section{Introduction}

Atom interferometers \cite{Kasevich1991,Cronin2009} have demonstrated excellent performance in measuring gravity \cite{Peters2001,Hu2013,Gillot2014,Abend2016}, gravity gradients \cite{McGuirk2002,Yu2006,Rosi2014} and rotations \cite{Gustavson2000,Durfee2006,Dutta2016}, using atoms in ballistic flight. In spite of being less well developped, trapped atom interferometers, for example using atom chips, \cite{Fortagh2002,Reichel2010}, would render the interrogation time independent of the atom's flight, permitting miniaturization and possibly longer measurement times. To date, atom-chip-based interferometers have been successfully demonstrated using Bose-Einstein condensates \cite{Schumm2005,Bohi2009}, but are subject to dephasing mechanisms resulting from atom-atom interactions \cite{Javanainen1997,Jo2007,Grond2010,Berrada2013}. Recently, we proposed a trapped atom interferometer using an ensemble of cold atoms above the Bose-Einstein condensation threshold (referred to as thermal atoms in the following) \cite{Ammar2014,DupontNivet2014}. This proposal is reminiscent of optical white light interferometry because of the necessity of keeping the path length difference between the two arms of the interferometer smaller than the coherence length. Similarly, the contrast decay of an atom chip interferometer using thermal atoms is related to the degree of asymmetry between the two arms \cite{DupontNivet2014}. Related contrast decay effects have been described in optical traps, for example in references \cite{Kuhr2003,Kuhr2005}, where the asymmetry results from state-dependent light shifts, and reference \cite{Hilico2015}, where the asymmetry is induced by spatial separation along the axis of a Gaussian beam.

In our experiment, the two arms of the interferometer correspond to two different internal states trapped in magnetic potentials, and we are able to control the effect of asymmetry without spatially separating the paths. The asymmetry can be tuned by adjusting the bias field, which results in slightly different magnetic moments as described by the Breit-Rabi formula. Using this technique in combination with evaporative cooling, we are able to measure the contrast decay time in an internal state interferometer (a Ramsey interferometer \cite{Ramsey1956}) for different values of the temperature and asymmetry, and compare our results with the theoretical predictions from reference \cite{DupontNivet2014}.

The focus of this paper is inhomogeneous dephasing, as manifested by the contrast decay of Ramey fringes \cite{Kuhr2005}. On the other hand, homogenous dephasing, caused for example by fluctuating magnetic fields, and probed by spin echo measurements \cite{Kuhr2005,Hahn1950}, is beyond the scope of this paper. We first give a brief review of the theoretical predictions in section \ref{sec_ConTime}. Then, we describe in section \ref{sec_exp} our experimental protocol and results. We finally compare the results to a simple model in section \ref{sec_dis}. We also discuss the identical spin rotation effect (ISRE) that was previously observed in similar experiments \cite{Lewandowski2002,Du2008,Du2009b,Deutsch2010,Kleine2011}.

\section{Theoretical model}
\label{sec_ConTime}

In this section, we briefly recall the simple model of reference \cite{DupontNivet2014} describing the influence of the asymmetry on the contrast decay time. We consider a Ramsey interferometer involving two internal states of the $^{87}$Rb ground state manifold, namely $\left|F=1,m_F=-1\right>\equiv\left|a\right>$ and $\left|F=2,m_F=1\right>\equiv\left|b\right>$, coupled by a two photon transition. During the whole interferometer sequence both states are maintained trapped, but not necessarily in identical potentials \cite{Bohi2009,Ammar2014}, leading to inhomogeneous dephasing. We suppose that the traps are harmonic but with slightly different frequencies along one of the trapping axes, namely $\omega_a=\omega$ for state $\left|a\right>$ and $\omega_b=\omega+\delta\omega$ for state $\left|b\right>$ with $|\delta\omega| \ll \omega$. We also assume that the gas is at sufficiently high temperature $T$ to be accurately described by a Boltzmann distribution. The relative asymmetry $|\delta\omega|/\omega$ then implies an upper bound on the contrast decay time $t_c$ given (up to a numerical factor on the order of unity) by \cite{DupontNivet2014}:
\begin{equation}
t_c \simeq \frac{\omega}{|\delta\omega|}\frac{\hbar}{k T}\;.
\label{eq_ContrastLaw}
\end{equation} 
Two differences between the experiment considered here and the model of reference \cite{DupontNivet2014} should be pointed out. First, in the model of reference \cite{DupontNivet2014}, the relative asymmetry was assumed to grow linearly from zero to some finite value $|\delta\omega|/\omega$ where it was held for some interrogation time, before being ramped back to zero. It was furthermore assumed that the ramp was slow enough that the initial population of the eigenstates was conserved throughout the sequence. By contrast, the splitting and recombination are very fast in our experiment (on the order of the $\pi/2$ pulse time). Still, we expect the populations to be approximately conserved (and the above formula to apply up to a numerical factor on the order of unity) as long as the relative asymmetry is small enough (typically smaller than $\hbar \omega/kT \gtrsim 10^{-3}$ in our experiment). Second, the model of reference \cite{DupontNivet2014} is one-dimensional, while in the experiment described here the asymmetry occurs along all three trapping axes (with identical relative asymmetry, as will be seen below). Again, we expect this not to change the results of \cite{DupontNivet2014} up to a numerical factor on the order of unity.

\section{Experiment}
\label{sec_exp}

\subsection{Tuning the asymmetry}

The two interferometer states $\left|a\right>\equiv\left|F=1,m_F=-1\right>$ and $\left|b\right>\equiv\left|F=2,m_F=1\right>$ are trapped by the same DC magnetic field. Because of the coupling between the nuclear angular momentum and the magnetic field, the magnetic moments (defined as the partial derivative of the energy with respect to the magnetic field) of the two states are slightly different, as described by the Breit-Rabi formula \cite{Steck2003}. As pointed out in reference \cite{Harber2002}, there is a ``magic" magnetic field $B_m^0 $ for which the effective magnetic moments of the two states $\left|a\right>$ and $\left|b\right>$ are identical. By changing the value of the field at the trap minimum around this value, we can go from a situation where the two traps are almost perfectly symmetric to a situation where they have significantly different frequencies.

Let us consider a static magnetic trapping field of the form $B(x)=B_0 + (\partial^2 B/\partial x^2) x^2$ where $\partial^2 B/\partial x^2 > 0$ and $x$ is a linear coordinate along one trapping axis. One can show \cite{DupontNivet2016}, expanding the Breit-Rabi formula \cite{Steck2003} up to the second order in the magnetic field, that the resulting trap frequencies are:
\begin{equation}
\omega_{a} \simeq \sqrt{\frac{\mu_Bg_J}{2m}\frac{\partial^2 B}{\partial x^2}}\sqrt{1+\frac{g_I}{g_J}\left(-5+4\frac{B_0}{B_m^0}\right)}
\end{equation}
for the state $\left|a\right>$, and:
\begin{equation}
\omega_{b} \simeq \sqrt{\frac{\mu_Bg_J}{2m}\frac{\partial^2 B}{\partial x^2}}\sqrt{1+\frac{g_I}{g_J}\left(3-4\frac{B_0}{B_m^0}\right)}
\end{equation}
for the state $\left|b\right>$, where the magic magnetic field reads \cite{Harber2002}:
\begin{equation}
B_m^0 \simeq - \frac{8g_IE_{hfs}}{3\mu_B(g_J-g_I)^2} \simeq 3.23~\mathrm{G} \;.
\end{equation}
In the above formulas, $m$ is the atomic mass, $\mu_B$ is the Bohr magneton, $g_J\simeq2.002$ and $g_I\simeq-9.95\cdot10^{-4}$ are the electron and nuclear spin $g$ factors \cite{Steck2003} respectively, and $E_{hfs}$ is the energy splitting between the two hyperfine ground states of $^{87}$Rb. The two different curvatures for the two traps result in a non-zero value for the asymmetry along each trapping axis:
\begin{equation}
\frac{\delta\omega}{\omega} = \frac{2|\omega_{b}-\omega_{a}|}{|\omega_{b}+\omega_{a}|} \simeq \frac{4}{1-g_I/(2g_J)} \left| \frac{g_I}{g_J} \left( 1 - \frac{B_0}{B_m^0} \right) \right|  \;.
\label{eq_AsymMod}
\end{equation}
As expected, the asymmetry depends on the magnetic field $B_0$ at the trap minimum and vanishes when the latter is equal to the magic field ($B_0=B_m^0$). The model can be easily generalized to 3 dimensions: the relative asymmetry is the same for all three trapping axes as can be seen in equation (\ref{eq_AsymMod}).

\begin{figure}
\centering  \includegraphics[width=0.65\textwidth]{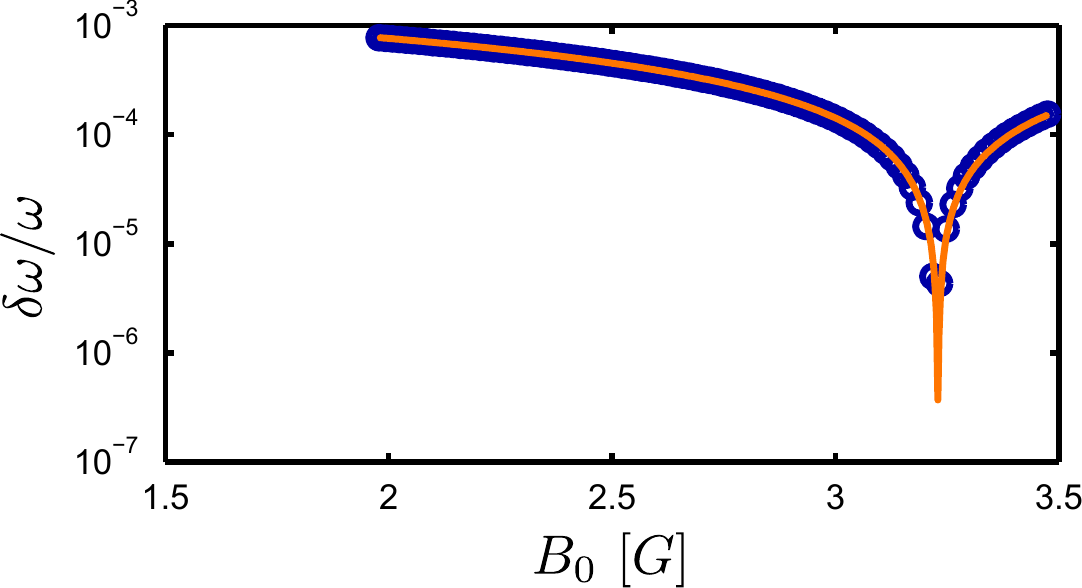}
\caption{\label{fig_Asym}  (Color online). Relative asymmetry along one axis as a function of the magnetic field at the trap minimum. The blue circles are the simulated values computed with our current distribution, and the orange solid line is the model of equation (\ref{eq_AsymMod}).  }
\end{figure}

On the chip, a dimple trap is created by two crossing wires and an external bias field. We change the trap field minimum by changing the magnitude and direction of this bias field. In the whole span of $B_0$ in figure \ref{fig_Asym}, the trap position changes by less than 1~$\mu$m and the trap frequencies, which are $\sim$85~Hz and $\sim$148~Hz in the horizontal plane and $\sim$161~Hz in the vertical plane, by less than 5~Hz (numbers are given for state $\left|b\right>$).

A simulation of our magnetic field geometry (performed using the known wire geometry on our atom chip) allows us to compute the potentials for the two states $\left|a\right>$ and $\left|b\right>$ and thus the asymmetry. In figure \ref{fig_Asym}, the asymmetry is shown as a function of the value of the magnetic field at the trap minimum.
\begin{table}
\begin{center}
\begin{tabular}{|c|c|c|}
\hline
 Trap minimum $B_0$ & Simulated asymmetry $\left.\frac{\delta\omega}{\omega}\right|_s$ & Inferred asymmetry $\left.\frac{\delta\omega}{\omega}\right|_i$  \\
\hline
 $2.456\pm0,012$~G & $(4.7\pm0.5)\cdot10^{-4}$ & $(3.3\pm0.3)\cdot10^{-4}$ \\
 $2.859\pm0,006$~G & $(2.3\pm0.3)\cdot10^{-4}$ & $(9.2\pm1.5)\cdot10^{-5}$ \\
 $3.264\pm0,009$~G & $(2.3\pm0.3)\cdot10^{-5}$ & $(3.3\pm0.4)\cdot10^{-5}$ \\
\hline
\end{tabular}
\end{center}
\caption{\label{tab_symmetrie} Trap asymmetry for different magnetic field minima $B_0$. We show the calculated values and those inferred from the contrast decay time curves of figure \ref{fig_CoherenceTime}. }
\end{table}

\begin{figure*}
\centering  \includegraphics[width=1\textwidth]{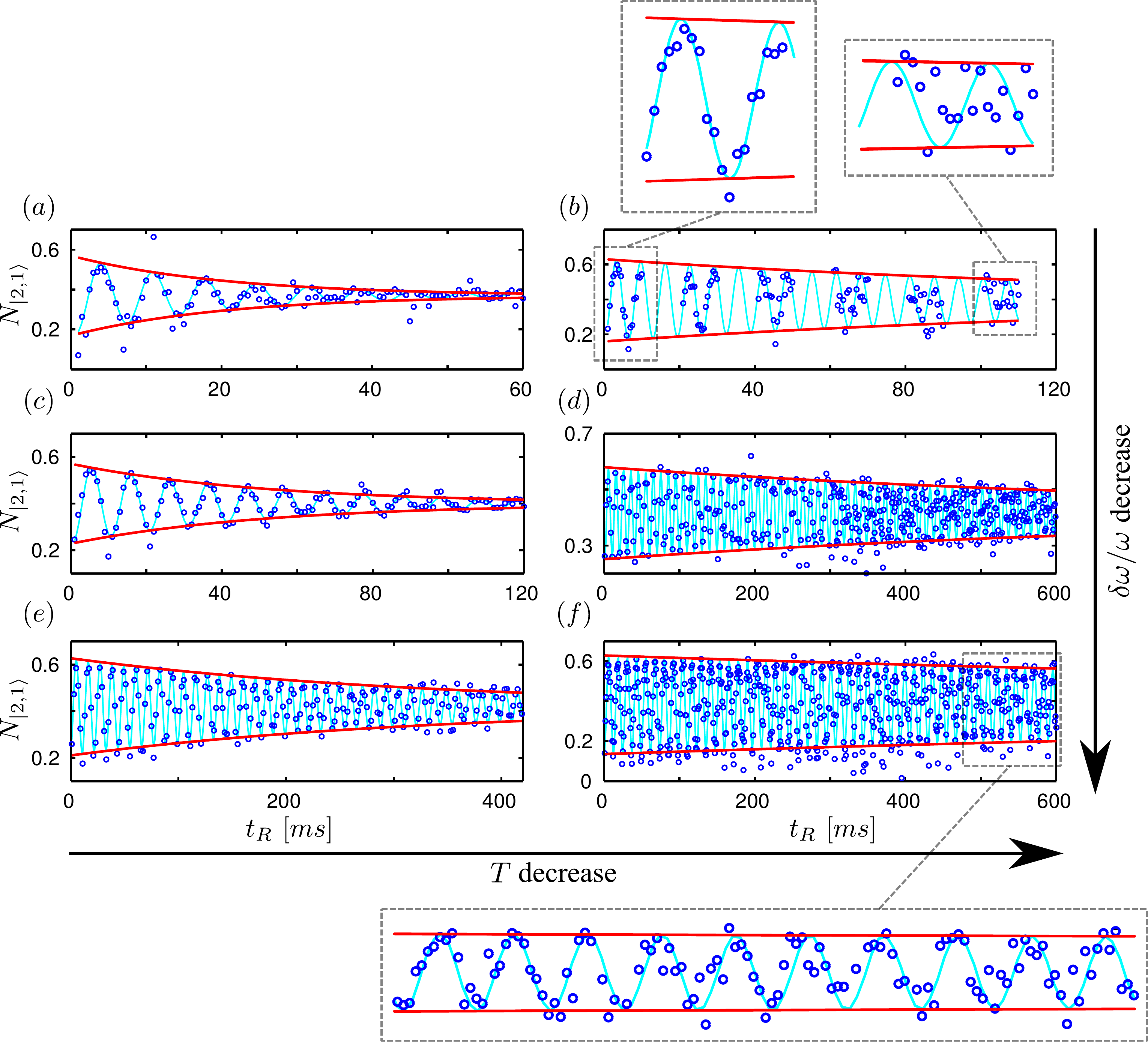}
\caption{\label{fig_RamseyTime}  (Color online). Ramsey fringes as a function of the interrogation time showing the behaviour of the contrast as a function of the cloud temperature and the asymmetry. Note that the horizontal scale (interrogation time $t_R$) is different in each plot. (a), (c) and (e) correspond to an atom cloud at a temperature of $\sim$500~nK, while (b), (d) and (f) correspond to $\sim$150~nK. For (a) and (b), $B_0=2.456$~G, corresponding to $\left.\frac{\delta\omega}{\omega}\right|_s=4.7\cdot10^{-4}$. For (c) and (d), $B_0=2.859$~G, corresponding to $\left.\frac{\delta\omega}{\omega}\right|_s=2.3\cdot10^{-4}$. For (e) and (f), $B_0=3.264$~G, corresponding to $\left.\frac{\delta\omega}{\omega}\right|_s=2.3\cdot10^{-5}$. The open blue circles stand for experimental data, the solid light blue lines are the fit of the Ramsey fringes and the solid red lines are the envelope of the Ramsey fringes. The insets show that for fields far from the magic field, the phase coherence is lost more rapidly than the contrast (see text).}
\end{figure*}

\begin{figure*}
\centering  \includegraphics[width=1\textwidth]{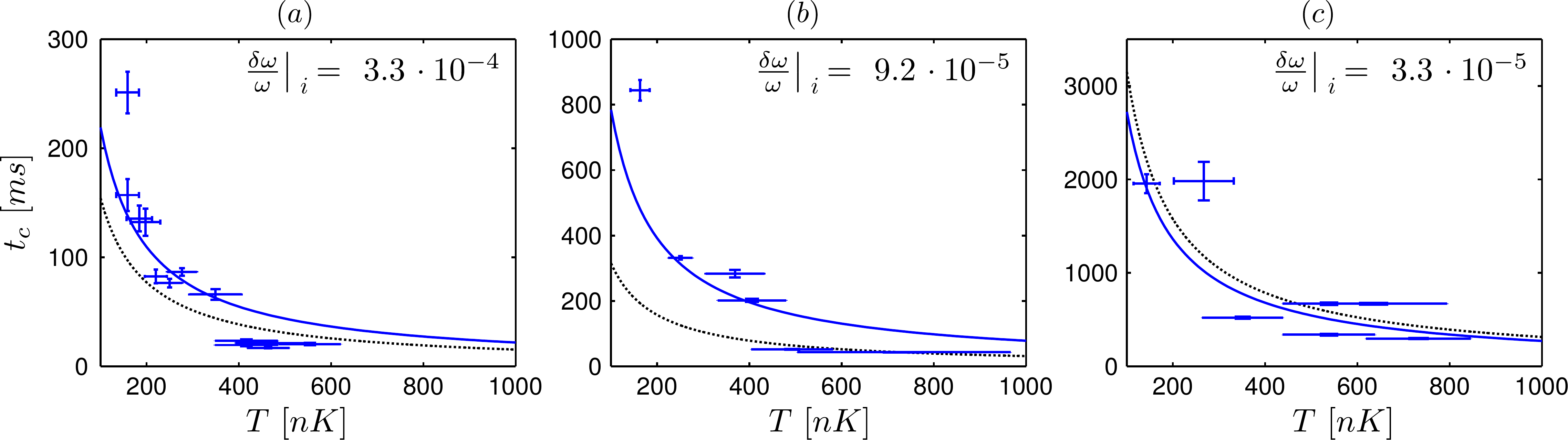}
\caption{\label{fig_CoherenceTime}  (Color online). Contrast decay time as a function of the atomic cloud temperature for three different values of the (inferred) asymmetry. The vertical scale (contrast decay time) is different in each plot. (a) $B_0=2.456$~G, corresponding to $\left.\frac{\delta\omega}{\omega}\right|_s=4.7\cdot10^{-4}$, (b) $B_0=2.859$~G, corresponding to $\left.\frac{\delta\omega}{\omega}\right|_s=2.3\cdot10^{-4}$ and (c) $B_0=3.264$~G, corresponding to $\left.\frac{\delta\omega}{\omega}\right|_s=2.3\cdot10^{-5}$. The crosses stand for the experimental data with error bars, the solid blue lines are the fit of the data (inferred values and errors are given in table \ref{tab_symmetrie}, column ``inferred asymmetry"). The dashed black lines are the model with the simulated values of the asymmetry. }
\end{figure*} 

\subsection{Experimental protocol and data}

We use the experimental setup described in references \cite{DupontNivet2016,DupontNivet2015} to trap a cloud of a few tens of thousands of $^{87}$Rb atoms in state $\left|F=2,m_F=2\right>$ on an atom chip, using forced radio-frequency evaporation to control the final temperature, which is typically chosen between 150~nK and 800~nK (with the Bose-Einstein condensation threshold around 110~nK temperature is given for state $\left|b\right>$). After evaporative cooling in state $\left|2,2\right>$, the cloud is transferred to state $\left|b\right>$ by microwave-stimulated Raman adiabatic passage as described in reference \cite{DupontNivet2015}. We then perform Ramsey spectroscopy between $\left|a\right>$ and $\left|b\right>$ for different values of the temperature and the asymmetry.

As described in the previous section, the bias field of the magnetic trap is tuned to three different values to investigate different values of the asymmetry. For these three biais fields, the magnetic field at the trap minimum is measured (see table \ref{tab_symmetrie}) and used to calculate the asymmetry using the simulation shown in figure \ref{fig_Asym} (values are given in the column labeled ``simulated asymmetry" $\left.\frac{\delta\omega}{\omega}\right|_s$ in table \ref{tab_symmetrie}). For each of these three values of the asymmetry, the Ramsey signal is recorded for different values of the temperature. The local oscillator used for the excitation has a short term stability below $5\cdot 10^{-12}$ at one second and long term drifts are corrected using a GPS clock signal. Examples of measured signals are shown in figure \ref{fig_RamseyTime} for two different values of the temperature. The two outputs of the interferometer are used to normalize the atom number in state $\left|b\right>$ by the total atom number \cite{Santarelli1999} (although only one is shown in figure \ref{fig_RamseyTime}). The interrogation time is limited to 600~ms so as not to overheat the atom chip. The envelopes of the fringes are extracted using a Hilbert transform \cite{Larkin1996} and then adjusted by a function of the form $\exp\left(-t/t_c\right)$ to infer the value of $t_c$ in equation (\ref{eq_ContrastLaw}) assuming asymmetry is the main source of contrast decay. The contrast decay time is plotted for the three values of the asymmetry as a function of the temperature in figure \ref{fig_CoherenceTime}. Each of these curves is adjusted by a function of the form of equation (\ref{eq_ContrastLaw}) with $\delta\omega/\omega$ the fitting parameter. The resulting fitted values are given in table \ref{tab_symmetrie} in the column ``inferred asymmetry" $\left.\frac{\delta\omega}{\omega}\right|_i$. 

\section{Discussion}
\label{sec_dis}

Our data is in reasonably good agreement with the model of equation (\ref{eq_ContrastLaw}), as can be seen in table \ref{tab_symmetrie}. The discrepancy can be possibly explained by the missing numerical factor in equation (\ref{eq_ContrastLaw}). On the other hand, closer inspection of figure \ref{fig_CoherenceTime} indicates that the data points seem to fall systematically above the fit for low temperature and below it for high temperature. This effect may point to trap anharmonicities which have been neglected up to this point. We can formulate a simple model of trap anharmonicity by considering the Hamiltonian:
\begin{equation}
\widehat{H}_i = \frac{\widehat{p}^2}{2m}+\frac{m\omega_i^2}{2}\widehat{x}^2+\frac{2}{\sqrt{15}}\sigma_i\hbar\omega_i\left(\frac{\hbar}{m\omega_i}\right)^{3/2}\widehat{x}^3 \;,
\end{equation}
where $i$ labels the internal atomic state $\left|i\right>$, $\sigma_i \ll 1$ is a dimensionless parameter quantifying the anharmonicity and $\widehat{p}$ (respectively $\widehat{x}$) is the momentum (respectively position) operator. The contrast decay time for such a potential is calculated in \cite{DupontNivet2014}, leading to:
\begin{equation}
\frac{1}{t_c^2} \sim 20 (D\sigma)^2\left(\frac{kT}{\hbar\omega}\right)^4 - 8 D\sigma\delta\omega\left(\frac{kT}{\hbar\omega}\right)^3 + \frac{3(\delta\omega)^2+2 (D\sigma)^2}{3}\left(\frac{kT}{\hbar\omega}\right)^2 \;,
\end{equation}
where $D\sigma = \omega\delta\sigma^2+\sigma^2\delta\omega$ with $\sigma^2=(\sigma_b^2+\sigma_a^2)/2$ and $\delta\sigma^2=\sigma_b^2-\sigma_a^2$, $\omega$ and $\delta\omega$ are defined in section \ref{sec_ConTime}. Perfectly harmonic potentials correspond to $D\sigma =0$ and we recover equation (\ref{eq_ContrastLaw}). A non vanishing $D \sigma$ will tend to accelerate the contrast decay at high temperature. It was not possible to adjust the data to the above model because of an insufficient signal to noise ratio. However we can estimate an order of magnitude for the maximum value of the $\sigma$ terms. For a temperature of 700~nK, $kT/(\hbar\omega)\sim$~120 (for $\omega$ we took the geometric mean of the trapping frequencies). Keeping only the first term, we find $D\sigma/\omega = \delta\sigma^2+\sigma^2(\delta\omega/\omega) \lesssim (\hbar\omega)^2/(\sqrt{20}\omega t_c (kT)^2 )$. In the most symmetric case (figure \ref{fig_CoherenceTime}.c) we observe a contrast decay time around 500~ms for temperature around 700~nK, thus $D\sigma/\omega \lesssim 3\cdot 10^{-8}$. This gives upper limits for the anharmonicity of the potentials: $\sigma  \lesssim 10^{-1}$, and for the difference between the cubic terms of the potentials: $\delta\sigma^2 \lesssim 3\cdot 10^{-8}$.

It can be noticed from figure \ref{fig_RamseyTime} that for some values of the asymmetry the phase coherence is lost even though the contrast remains appreciable (see for example subfigure b). We attribute this to magnetic field fluctuations (estimated to $\delta B \simeq 5$~mG RMS in our experiment \cite{DupontNivet2016}), whose effect on coherence becomes significant as we move away from the magic magnetic field \cite{Treutlein2006}. 

\begin{figure*}
\centering  \includegraphics[width=1\textwidth]{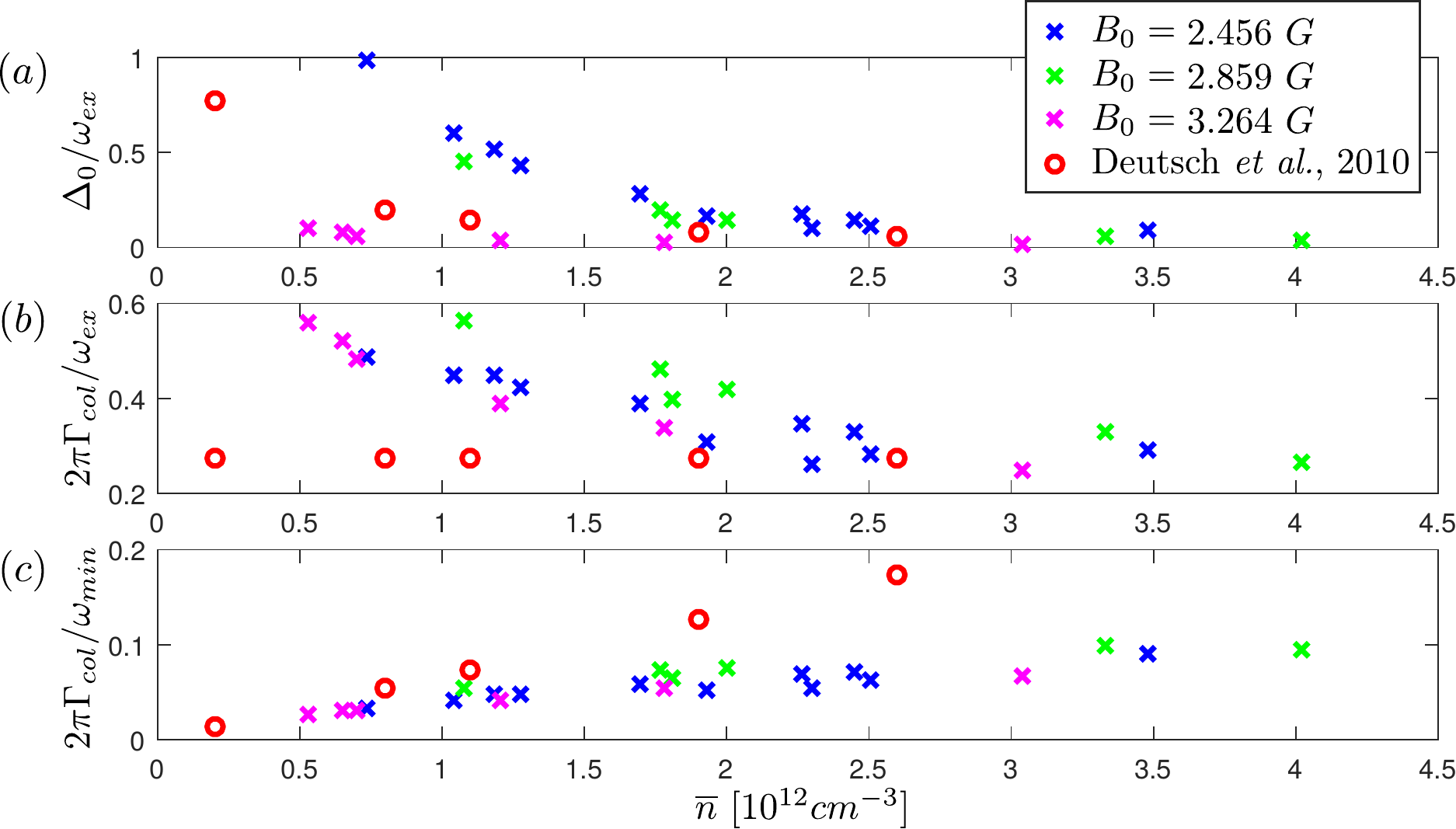}
\caption{\label{fig_ISREcond}  (Color online). The three relevant dimensionless parameters to assess the existence of ISRE. (a) $\Delta_0/\omega_{ex}$, (b) $2\pi\Gamma_{col}/\omega_{ex}$ and (c) $2\pi\Gamma_{col}/\omega_{min}$ as a function of the mean atomic density $\overline{n}$ in unit of $10^{12}$~cm$^{-3}$. The three sets of crosses stand for our experimental data, while the open circles stand for the data of reference \cite{Deutsch2010} where ISRE was observed. }
\end{figure*} 

Our experimental conditions (atomic density, temperature and inhomogeneity) are close to those of reference \cite{Deutsch2010} where an identical spin rotation effect (ISRE) results in long contrast decay times ($\sim60$~s) and contrast revivals. One might wonder why this effect is not visible in our experiment. The ISRE regime is characterized by three time scales \cite{Deutsch2010}: i) the inhomogeneity $\Delta_0=|kT\delta\omega/\hbar\omega - \gamma\overline{n}/4|$ with $\overline{n}$ the mean atomic density and $\gamma=4\pi\hbar(a_{bb}-a_{aa})/m$ where $a_{ij}$ is the scattering length between states $\left|i\right>$ and $\left|j\right>$, ii) the elastic collision rate $\Gamma_{col}=(32/3)a_{ab}^2\overline{n}(\pi k T /m)^{1/2}$ and iii) the exchange energy $\omega_{ex}=4\pi\hbar |a_{ab}| \overline{n}/m$. For ISRE to occur, the following three conditions must be satisfied \cite{Deutsch2010}: i) $\Delta_0/\omega_{ex} \ll 1$, ii) $2\pi\Gamma_{col}/\omega_{ex} \ll 1$ and iii) $2\pi\Gamma_{col}/\omega_{min} \ll 1$ where $\omega_{min}$ is the smallest trapping frequency. To get a more quantitative criterion, we compare our experimental values of these parameters, for various densities, to the numbers from reference \cite{Deutsch2010} where ISRE was observed (figure \ref{fig_ISREcond}). As can be seen in this figure, there are some situations (for example $B_0=3.264$~G and $\overline{n} > 2\cdot10^{12}$~cm$^3$) where all three relevant quantities are smaller in our case than in reference \cite{Deutsch2010}. Yet, we do not see ISRE in these cases. 

In an attempt to reconcile our observations with those of \cite{Deutsch2010}, we developed a model whose details will be published elsewhere \cite{DupontNivet2017}. This model indicates that the trap anisotropy could significantly enhance ISRE. The identical spin rotation effect arises from forward collisions between cold atoms which tend to be situated at the bottom of the trap and hot atoms which tend to sample magnetic fields far from the minimum. In the case of a cigar shaped trap, a hot atom oscillating in the trap comes close to the minimum of the trap at each oscillation enhancing forward collisions with cold atoms. In the case of an isotropic trap, a hot atom can oscillate without coming close to the trap minimum thus reducing the number of forward collisions with cold atoms and also the identical spin rotation effect. Our trapping potential is closer to an isotropic trap (frequencies 85~Hz, 148~Hz, 161~Hz) than that in reference \cite{Deutsch2010} (32~Hz, 97~Hz, 121~Hz), which might explain the discrepancy.

\section{Conclusion}

We have shown that the simple model derived in reference \cite{DupontNivet2014} describes reasonably well the contrast decay in our experiment, confirming the important role of symmetry in atom interferometry with thermal atoms. The next step is to introduce a spatial separation between the two internal states, for example  by using near-field microwave gradients \cite{Bohi2009,Ammar2014,DupontNivet2014}. In this context, equation (\ref{eq_ContrastLaw}) could serve as a benchmark to assess the required degree of symmetry in the design of future atom-chip interferometers.

\ack
This work has been carried out within the OnACIS project ANR-13-ASTR-0031 funded by the French National Research Agency (ANR) in the frame of its 2013 Astrid program. S.S. acknowledges funding from the European Union under the Marie Sklodowska Curie Individual Fellowship Programme H2020-MSCA-IF-2014 (project number 658253).

\section*{References}
\bibliographystyle{unsrt}
\bibliography{biblio}

\providecommand{\noopsort}[1]{}\providecommand{\singleletter}[1]{#1}%
\begin{thebibliography}{10}

\bibitem{Kasevich1991}
M.~Kasevich and S.~Chu.
\newblock Atomic interferometry using stimulated {R}aman transitions.
\newblock {\em Phys. Rev. Lett.}, 67:181--184, Jul 1991.

\bibitem{Cronin2009}
A.~Cronin, J.~Schmiedmayer, and D.~Pritchard.
\newblock Optics and interferometry with atoms and molecules.
\newblock {\em Rev. Mod. Phys.}, 81:1051--1129, Jul 2009.

\bibitem{Peters2001}
A.~Peters, K.~Y. Chung, and S.~Chu.
\newblock High-precision gravity measurements using atom interferometry.
\newblock {\em Metrologia}, 38(1):25, 2001.

\bibitem{Hu2013}
Z.-K. Hu, B.-L. Sun, X.-C. Duan, M.-K. Zhou, L.-L. Chen, S.~Zhan, Q.-Z. Zhang,
  and J.~Luo.
\newblock Demonstration of an ultrahigh-sensitivity atom-interferometry
  absolute gravimeter.
\newblock {\em Phys. Rev. A}, 88(4):043610, 2013.

\bibitem{Gillot2014}
P.~Gillot, O.~Francis, A.~Landragin, F.~P. Dos~Santos, and S.~Merlet.
\newblock Stability comparison of two absolute gravimeters: optical versus
  atomic interferometers.
\newblock {\em Metrologia}, 51(5):L15, 2014.

\bibitem{Abend2016}
S.~Abend, M.~Gebbe, M.~Gersemann, H.~Ahlers, H.~M{\"u}ntinga, E.~Giese,
  N.~Gaaloul, C.~Schubert, C.~L{\"a}mmerzahl, W.~Ertmer, W.~P. Schleich, and
  E.~M. Rasel.
\newblock Atom-chip fountain gravimeter.
\newblock {\em Phys. Rev. Lett.}, 117(20):203003, 2016.

\bibitem{McGuirk2002}
J.~McGuirk, G.~Foster, J.~Fixler, M.~Snadden, and M.~Kasevich.
\newblock Sensitive absolute-gravity gradiometry using atom interferometry.
\newblock {\em Phys. Rev. A}, 65:033608, Feb 2002.

\bibitem{Yu2006}
N.~Yu, J.~M. Kohel, J.~R. Kellogg, and L.~Maleki.
\newblock Development of an atom-interferometer gravity gradiometer for gravity
  measurement from space.
\newblock {\em Appl. Phys. B}, 84(4):647--652, 2006.

\bibitem{Rosi2014}
G.~Rosi, F.~Sorrentino, L.~Cacciapuoti, M.~Prevedelli, and G.~M. Tino.
\newblock Precision measurement of the newtonian gravitational constant using
  cold atoms.
\newblock {\em Nature}, 510(7506):518--521, 2014.

\bibitem{Gustavson2000}
T.~Gustavson, A.~Landragin, and M.~Kasevich.
\newblock Rotation sensing with a dual atom-interferometer sagnac gyroscope.
\newblock {\em Classical Quant. Grav.}, 17(12):2385, 2000.

\bibitem{Durfee2006}
D.~S. Durfee, Y.~K. Shaham, and M.~A. Kasevich.
\newblock Long-term stability of an area-reversible atom-interferometer sagnac
  gyroscope.
\newblock {\em Phys. Rev. Lett.}, 97:240801, Dec 2006.

\bibitem{Dutta2016}
I.~Dutta, D.~Savoie, B.~Fang, B.~Venon, C.~L.~Garrido Alzar, R.~Geiger, and
  A.~Landragin.
\newblock Continuous cold-atom inertial sensor with 1 nrad/sec rotation
  stability.
\newblock {\em Phys. Rev. Lett.}, 116(18):183003, 2016.

\bibitem{Fortagh2002}
J.~Fort\'agh, H.~Ott, S.~Kraft, A.~G\"unther, and C.~Zimmermann.
\newblock Surface effects in magnetic microtraps.
\newblock {\em Phys. Rev. A}, 66:041604, Oct 2002.

\bibitem{Reichel2010}
J.~Reichel and V.~Vuletic.
\newblock {\em Atom Chips}.
\newblock John Wiley \& Sons, 2010.

\bibitem{Schumm2005}
T.~Schumm, S.~Hofferberth, L.~M. Andersson, S.~Wildermuth, S.~Groth,
  I.~Bar-Joseph, J.~Schmiedmayer, and P.~Kruger.
\newblock Matter-wave interferometry in a double well on an atom chip.
\newblock {\em Nat. Phys.}, 1:57--62, 2005.

\bibitem{Bohi2009}
P.~B{\"o}hi, M.~Riedel, J.~Hoffrogge, J.~Reichel, T.~Hansch, and P.~Treutlein.
\newblock Coherent manipulation of bose-einstein condensates with
  state-dependent microwave potentials on an atom chip.
\newblock {\em Nat. Phys.}, 5:592--597, 2009.

\bibitem{Javanainen1997}
J.~Javanainen and M.~Wilkens.
\newblock Phase and phase diffusion of a split bose-einstein condensate.
\newblock {\em Phys. Rev. Lett.}, 78:4675--4678, Jun 1997.

\bibitem{Jo2007}
G.-B. Jo, Y.~Shin, S.~Will, T.~A. Pasquini, M.~Saba, W.~Ketterle, D.~E.
  Pritchard, M.~Vengalattore, and M.~Prentiss.
\newblock Long phase coherence time and number squeezing of two bose-einstein
  condensates on an atom chip.
\newblock {\em Phys. Rev. Lett.}, 98:030407, Jan 2007.

\bibitem{Grond2010}
J.~Grond, U.~Hohenester, I.~Mazets, and J.~Schmiedmayer.
\newblock Atom interferometry with trapped bose-einstein condensates: impact of
  atom-atom interactions.
\newblock {\em New J. Phys.}, 12(6):065036, 2010.

\bibitem{Berrada2013}
T.~Berrada, S.~van Frank, R.~B{\"u}cker, T.~Schumm, J.-F. Schaff, and
  J.~Schmiedmayer.
\newblock Integrated {M}ach--{Z}ehnder interferometer for {B}ose--{E}instein
  condensates.
\newblock {\em Nat. Commun.}, 4, 2013.

\bibitem{Ammar2014}
M.~Ammar, M.~Dupont-Nivet, L.~Huet, J.-P. Pocholle, P.~Rosenbusch,
  I.~Bouchoule, C.~I. Westbrook, J.~Est\`eve, J.~Reichel, C.~Guerlin, and
  S.~Schwartz.
\newblock Symmetric microwave potentials for interferometry with thermal atoms
  on a chip.
\newblock {\em Phys. Rev. A}, 91:053623, May 2015.

\bibitem{DupontNivet2014}
M.~Dupont-Nivet, C.~I. Westbrook, and S.~Schwartz.
\newblock Contrast and phase-shift of a trapped atom interferometer using a
  thermal ensemble with internal state labelling.
\newblock {\em New J. Phys.}, 18(11):113012, 2016.

\bibitem{Kuhr2003}
S.~Kuhr, W.~Alt, D.~Schrader, I.~Dotsenko, Y.~Miroshnychenko, W.~Rosenfeld,
  M.~Khudaverdyan, V.~Gomer, A.~Rauschenbeutel, and D.~Meschede.
\newblock Coherence properties and quantum state transportation in an optical
  conveyor belt.
\newblock {\em Phys. Rev. Lett.}, 91(21):213002, 2003.

\bibitem{Kuhr2005}
S.~Kuhr, W.~Alt, D.~Schrader, I.~Dotsenko, Y.~Miroshnychenko,
  A.~Rauschenbeutel, and D.~Meschede.
\newblock Analysis of dephasing mechanisms in a standing-wave dipole trap.
\newblock {\em Phys. Rev. A}, 72(2):023406, 2005.

\bibitem{Hilico2015}
A.~Hilico, C.~Solaro, M.-K. Zhou, M.~Lopez, and F.~Pereira~dos Santos.
\newblock Contrast decay in a trapped-atom interferometer.
\newblock {\em Phys. Rev. A}, 91(5):053616, 2015.

\bibitem{Ramsey1956}
N.~Ramsey.
\newblock {\em Molecular beams}.
\newblock Oxford University Press, 1956.

\bibitem{Hahn1950}
E.~L. Hahn.
\newblock Spin echoes.
\newblock {\em Phys. Rev.}, 80(4):580, 1950.

\bibitem{Lewandowski2002}
H.~J. Lewandowski, D.~M. Harber, D.~L. Whitaker, and E.~A. Cornell.
\newblock Observation of anomalous spin-state segregation in a trapped
  ultracold vapor.
\newblock {\em Phys. Rev. Lett.}, 88:070403, Jan 2002.

\bibitem{Du2008}
X.~Du, L.~Luo, B.~Clancy, and J.~E. Thomas.
\newblock Observation of anomalous spin segregation in a trapped fermi gas.
\newblock {\em Phys. Rev. Lett.}, 101:150401, Oct 2008.

\bibitem{Du2009b}
X.~Du, Y.~Zhang, J.~Petricka, and J.~E. Thomas.
\newblock Controlling spin current in a trapped fermi gas.
\newblock {\em Phys. Rev. Lett.}, 103:010401, Jul 2009.

\bibitem{Deutsch2010}
C.~Deutsch, F.~Ramirez-Martinez, C.~Lacro\^ute, F.~Reinhard, T.~Schneider,
  J.~N. Fuchs, F.~Pi\'echon, F.~Lalo\"e, J.~Reichel, and P.~Rosenbusch.
\newblock Spin self-rephasing and very long coherence times in a trapped atomic
  ensemble.
\newblock {\em Phys. Rev. Lett.}, 105:020401, Jul 2010.

\bibitem{Kleine2011}
G.~Kleine~B\"uning, J.~Will, W.~Ertmer, E.~Rasel, J.~Arlt, C.~Klempt,
  F.~Ramirez-Martinez, F.~Pi\'echon, and P.~Rosenbusch.
\newblock Extended coherence time on the clock transition of optically trapped
  rubidium.
\newblock {\em Phys. Rev. Lett.}, 106:240801, Jun 2011.

\bibitem{Steck2003}
D.~A. Steck.
\newblock Rubidium 87 {D} line data, revision 1.6.
\newblock {\em Source--http://steck.us/alkalidata}, 2003.

\bibitem{Harber2002}
D.~M. Harber, H.~J. Lewandowski, J.~M. McGuirk, and E.~A. Cornell.
\newblock Effect of cold collisions on spin coherence and resonance shifts in a
  magnetically trapped ultracold gas.
\newblock {\em Phys. Rev. A}, 66:053616, Nov 2002.

\bibitem{DupontNivet2016}
M.~Dupont-Nivet.
\newblock {\em Vers un acc\'el\'erom\'etre atomique sur puce}.
\newblock PhD thesis, Universit\'e Paris Saclay, 2016.

\bibitem{DupontNivet2015}
M.~Dupont-Nivet, M.~Casiulis, T.~Laudat, C.~I. Westbrook, and S.~Schwartz.
\newblock Microwave-stimulated raman adiabatic passage in a bose-einstein
  condensate on an atom chip.
\newblock {\em Phys. Rev. A}, 91:053420, May 2015.

\bibitem{Santarelli1999}
G.~Santarelli, Ph. Laurent, P.~Lemonde, A.~Clairon, A.~G. Mann, S.~Chang, A.~N.
  Luiten, and C.~Salomon.
\newblock Quantum projection noise in an atomic fountain: A high stability
  cesium frequency standard.
\newblock {\em Phys. Rev. Lett.}, 82:4619--4622, Jun 1999.

\bibitem{Larkin1996}
K.~G. Larkin.
\newblock Efficient nonlinear algorithm for envelope detection in white light
  interferometry.
\newblock {\em J. Opt. Soc. Am. A}, 13(4):832--843, 1996.

\bibitem{Treutlein2006}
P.~Treutlein, T.~W. H\"ansch, J.~Reichel, A.~Negretti, M.~A. Cirone, and
  T.~Calarco.
\newblock Microwave potentials and optimal control for robust quantum gates on
  an atom chip.
\newblock {\em Phys. Rev. A}, 74:022312, Aug 2006.

\bibitem{DupontNivet2017}
M.~Dupont-Nivet, S.~Schwartz, and C.~I. Westbrook.
\newblock Effect of trap symmetry and atom-atom interactions on the contrast
  decay and the phase-shift of a double well trapped ramsey interferometer with
  internal states labelling.
\newblock {\em In preparation}, 2017.

\end{thebibliography}

\end{document}